# An Urban Multi-Operator QoE-Aware Dataset for Cellular Networks in Dense Environments

Muhammad Kabeer[1,2], Rosdiadee Nordin[1], Mehran Behjati[1], Farah Yasmin binti Mohd Shaharuddin[3]

1. Faculty of Engineering and Technology, Sunway University, No. 5, Jalan Universiti, Bandar Sunway 47500, Selangor, Malaysia

2. Department of Computer Science, Federal University Dutsinma, Katsina Nigeria

3. Faculty of Engineering and Built Environment, Universiti Kebangsaan Malaysia, Bangi 43600, Selangor, Malaysia

**Graphical Abstract**

**Abstract**

Urban cellular networks face complex performance challenges due to high infrastructure density, varied user mobility, and diverse service demands. While several datasets address network behaviour across different environments, there is a lack of datasets that captures user-centric Quality of Experience (QoE), and diverse mobility patterns needed for efficient network planning and optimization solutions, which are important for QoE-driven optimizations and mobility management. This study presents a curated dataset of 30,925 labelled records, collected using *GNetTrack Pro* within a 2 km² dense urban area, spanning three major commercial network operators. The dataset captures key signal quality parameters (e.g., RSRP, RSRQ, SNR), across multiple real-world mobility modes including pedestrian routes, canopy walkways, shuttle buses, and Bus Rapid Transit (BRT) routes. It also includes diverse network traffic scenarios including (1) FTP upload/download, (2) video streaming, and (3) HTTP browsing. A total of 132 physical cell sites were identified and validated through OpenCellID and on-site field inspections, illustrating the high cell density characteristic of 5G and emerging heterogeneous network deployment. The dataset is particularly suited for machine learning applications, such as handover optimization, signal quality prediction, and multi-operator performance evaluation. Released in a structured CSV format with accompanying preprocessing and visualization scripts, this dataset offers a reproducible, application-ready resource for researchers and practitioners working on urban cellular network planning and optimization.

**Specification Table**

| Field | Description |
|---|---|
| Subject | Computer Science, Telecommunications, Data Science |
| Specific subject area | Mobile Network Performance, 4G/5G Signal Analysis, Mobility-aware Cellular Coverage Evaluation |
| Type of data | Raw, Cleaned |
| Data Collection | The data were acquired using an Android-based data collection application *GNetTrack Pro* developed by Gyokov Solutions. The application logged radio parameters including signal strength metrics, location, network technology, and device mobility in real time. Different mobility patterns walk, busses etc are used and diverse user-context like FTP, 1080p Video streaming and HTTP are considered to capture the most realistic user experience. |
| Data format | CSV (Comma-Separated Values) |
| Description of data collected | Data were collected within a radius of approximately 2 km around Sunway University, Selangor, Malaysia. The measurements distribution is around 70% 5G and 30% 4G showing massive deployment of 5G in the area. Operators were anonymized during preprocessing and data were handled to remove duplicates, and outliers as per 3GPP TS 36.214 standard in important columns resulting in 30,925 with balanced distribution among the operators. Altitude readings range from -32 meters to 73 meters, spanning a range of 105 metres with a mean of 27.42 meters demonstrating the unique characteristics of the area that has not been seen in similar datasets. Data were collected across three anonymized mobile network operators labelled Operator A, B, and C. |
| Data source location | Institution: Sunway University, Selangor<br>Country: Malaysia. |
| Data accessibility | Repository name: Mendeley Data<br>Data identification number DOI: 10.17632/dx5xyyfz2y.1<br>URL: https://data.mendeley.com/datasets/dx5xyyfz2y/1 |
| Related Research Article | None |

**1.0 Value of the Data**

- This dataset captures cellular network performance in a dense urban environment, incorporating 46 radio parameters from both serving and neighbouring cells, along withmobile user-context metrics, offering a unique combination not commonly found in existing datasets.

- It includes diverse and realistic mobility scenarios such as pedestrian routes, canopy walkways, shuttle buses, and Bus Rapid Transit (BRT) lines typical of urban areas, providing valuable insights into signal behaviour under diverse movement conditions.

- Traffic-aware profiling across HTTP browsing, FTP transfers, and 1080p Video Streaming allows researchers to assess QoE in conjunction with conventional signal metrics.

- Empirical validation of 132 physical cell locations through OpenCellID and on-site verification supports research on small cell deployment strategies, urban infrastructure planning, and heterogeneous network design.

- The dataset is well-suited for a wide range of ML and deep learning tasks, such as signal metrics regression, context-aware handover management, classification, multi-operator performance analysis, and mobility-aware optimization, as shown in [1], [2], [3] [4].

## 2.0 Background

Urban cellular networks are facing increasingly complex performance challenges due to high infrastructure density, user mobility, and the growing diversity of service demands [3]. The evolution of these networks, from early voice-only systems to the current 5G era, has been rapid, driven primarily by escalating bandwidth requirements of modern applications such as social media, online gaming, and augmented/virtual reality, which have often exceeded initial capacity projections [5], [6].

To meet these rising demands, the fifth generation (5G) of cellular communication standards has been developed, offering substantially higher data rates, lower latency, and greater device connectivity [7]. However, maintaining these performance benefits in real-world deployments remains a significant challenge. To evaluate and improve the cellular network behaviour, a variety of datasets have been developed, many of which focus on 4G and 5G technologies. These datasets typically contain channel-related parameters (e.g., signal strength) and context information (e.g., geographical location, user mobility patterns) [2], [3].

However, with the emergence and massive deployment of 5G networks, there is a growing need for more holistic and user-centric datasets. Next-generation datasets should capture t the unique attributes of 5G networks, while also incorporating detailed user-related KPIs. Such richer data is crucial for optimising the QoE [8] The ability to predict network performance and proactively manage it under diverse conditions is essential for realising the full potential of 5G and beyond [5].

The prediction of key network performance metrics such as RSRP as in [1] and downlink throughput in [9] is increasingly driven by machine learning (ML) techniques. By leveraging diverse input features, including channel quality indicators and spatial context, ML models enable more accurate forecasting of network behaviour and support intelligent planning, optimization, and adaptation. Understanding the factors influencing these KPIs and developing robust predictive models is essential for maintaining high QoE in modern mobile networks. [1][9]Recent works have shown that ML-based tools, such as the Machine Learning Based Online Coverage Estimator (MLOE) using Random Forests, outperform traditional methods in RSRP prediction for terrestrial networks [10], while similar approaches enable reliable RSRP and RSRQ forecasting for aerial communications in drone networks [11]. More recently, a triple-layer machine learning model combining linear regression, bagged trees, and Gaussian process regression has demonstrated over 90% testing accuracy in predicting RSRP, RSRQ, and other KPIs for cellular-connected UAVs [12].

While a recent comprehensive cellular network dataset collected in Brazil [2] provided valuable insights from the Amazon region typically dominated by 4G and less dense deployment, there is still a noticeable gap in datasets tailored to the complex conditions of dense urban environments for understanding realistic handover behaviour and mitigating mobility challenges. To address this gap, it is essential to develop a multi-operator dataset that combines realistic mobility patterns (e.g., pedestrian routes, canopy walkways, shuttle buses, and BRT lines), traffic-aware profiling (e.g., HTTP, FTP, 1080p video streaming), and empirically validated physical cell locations.

Such a dataset would be highly suitable for advanced analytical methods, including ML and deep learning, and could support tasks such as signal quality regression, coverage prediction, and context-aware handover management [13], [14]. Focusing on the distinct challenges of dense urban areas, this work aims to provide novel and actionable insights for future cellular network research and development.

## 3.0 Data Description

This study presents a real-world dataset collected by the Wireless Research Laboratory at Sunway University to better understand mobile network performance in a dense urban environment. Data collection focused on the outdoor area around the university campus in Selangor, Malaysia, which includes a mix of open spaces, high-rise buildings, and transportation routes like shuttle services and BRT lines. These diverse surroundings enabled the capture of a rich and varied dataset reflecting real-world signal behaviour across different urban conditions.

The dataset comprises 30,925 post processing individual records, collected using *GNetTrack Pro*, an Android-based network logging tool. Each entry includes a range of signal parameters such as RSRP, RSRQ, SNR as well as context-aware features like location, altitude, network type, and user speed. Measurements were gathered across three commercial mobile network operators (anonymized for privacy), all providing 4G and 5G services . The collected values were validated against 3GPP standard, and any outliers were capped or removed to ensure data integrity.

To reflect real-world usage patterns, the dataset includes records collected under different mobility, such as walking on pedestrian pathways and canopy walkways, commuting by university shuttle buses, and traveling via the BRT system. For simplicity, movements were categorized based on speed into "walking" and "driving", with driving category accounting for a slightly more than half of the dataset. These mobility labels enrich the dataset and support research into mobility-aware network optimization techniques.

A rigorous data cleaning process was applied: rows with missing values in key columns were removed, duplicates entries were eliminated, and non-standard operator values mapped consistently. The data was collected within 2km radius of Sunway University, the measurements span across 132 physical cell nodes including shared nodes, the nodes were cross validated with crowdsourced OpencellID and site visits. It also includes altitude information, ranging from –32 meters to 73 meters above sea level, which can be helpful in studying signal propagation in vertical environments.

This dataset provides a robust foundation for a wide range of research applications. It supports machine learning tasks like classification and regression, mobility-based handover management, and fog-enabled Digital Twin based self-optimization of cellular networks research [15] [16] [17]. By combining detailed signal metrics with realistic movement patterns and dense urban context, it offers valuable insight into current and future mobile network planning.

## 4.0 Experimental Design, Materials and Methods

Several recent efforts have focused on developing datasets to capture 4G and 5G mobile network performance using real-world measurements [1], [2], [3] and synthetic data [18], with varying degrees of public accessibility. These resources are designed to help researchers analyse network behaviour at scale, particularly in urban environments where high user density, diverse mobility patterns, and diverse infrastructure conditions introduce significant challenges for maintaining consistent quality of service.

Building on these efforts, this work presents a real-world dataset collected entirely from live mobile network interactions in an urban area. Unlike emulated or simulator-based datasets, which are useful for theoretical modelling, our dataset emphasizes actual signal behaviour under dynamic conditions including pedestrian movement, public transport, and vehicle-based commuting within a compact urban area around Sunway University.

The data collection process was conducted independently of mobile network operators, using the same model of Android device and *GNetTrack Pro* application across operators to minimize device bias. This methodological consistency, combined with the diversity of recorded radio parameters and contextual features, allows for an objective study of network KPIs such as signal quality, coverage reliability, user-context and the influence of mobility on service degradation.

Real-time data collection in live 5G environments presents operational complexities, particularly due to the need for high temporal and spatial granularity. In this context, our dataset addresses a critical gap by offering structured, validated, and reproducible signal logs, suitable for exploratory data analysis as well as advanced applications such as machine learning-based performance prediction, QoE profiling, and context-aware mobility optimization.

This approach is in line with emerging research directions that emphasize the importance of open-access, field-validated datasets to advance cellular network optimization in real-world deployments [2].

Table 1 presents 46 extracted features collected using the GNetTrack Pro application, along with 5 additional engineered features designed to support advanced analyses. The 'Mobility' feature was derived from user speed and categorized into two simplified classes to facilitate classification tasks. 'Node_Longitude' and 'Node_Latitude' were estimated using the centroid of all associated measurement points to approximate the location of serving cells; these locations were subsequently validated through crowdsourced OpenCellID data and on-site inspections.

The 'Session' feature was introduced to help deep learning models identify the start and end of measurement sessions, preserving temporal dependencies critical for sequence-based predictions [19]. 'ElapsedTime' represents the duration of user equipment (UE) stay within a given cell before handover, simplifying the interpretation of timestamp sequences in mobility-related studies.

These engineered features, among others, can significantly enhance research applications. For example, using the UE's captured GPS coordinates and the estimated serving node locations, the distance between the user and the base station can be calculated for further spatial analysis.

**Table 1.** Data dictionary with relevant features

| Feature Name | Description | Data Type |
|---|---|---|
| Timestamp | Date and time of the measurement (e.g., formatted as YYYY-MM-DD HH:MM:SS). | Datetime |
| Longitude | Longitude of the device's location | Float |
| Latitude | Latitude of the device's location | Float |
| Speed | Device's speed during measurement for urban movement mostly less than 50km/h | Float |
| Operatorname | Name of the network operator A, B and C. | String |
| Node | Identifier for the serving base station (eNodeB/gNodeB) | String |
| CellID | Unique identifier for the serving cell | Integer |
| LAC | Location Area Code for network tracking | Integer |
| NetworkTech | Radio access technology of the serving cell, either 4G or 5G. | String |
| Level | Reference Signal Received Power (RSRP) of the serving cell, typically -140 to -44 dBm. | Float |
| Qual | Reference Signal Received Quality (RSRQ) of the serving cell, typically -19.5 to -3 dB. | Float |
| SNR | Signal-to-Noise Ratio of the serving cell, typically -20 to 30 dB. | Float |
| CQI | Channel Quality Indicator for scheduling, ranging from 0 to 15. | Integer |
| LTERSSI | Received Signal Strength Indicator for LTE, typically -120 to -30 dBm. | Float |
| ARFCN | Absolute Radio Frequency Channel Number for the serving cell | Integer |
| DL_bitrate | Downlink data rate for video streaming | Float |
| UL_bitrate | Uplink data rate for video streaming | Float |
| PSC | Physical Cell Identity (PCI) of the serving cell | Integer |
| Altitude | Device altitude above sea level | Float |
| Accuracy | GPS accuracy of the location measurement | Float |
| State | Network connection state | String |
| SERVINGTIME | Time spent connected to the serving cell in seconds | Float |
| BANDWIDTH | Channel bandwidth of the serving cell | Float |
| SecondCell_NODE | Identifier for the secondary base station | String |
| SecondCell_CELLID | Cell ID of the secondary cell | Integer |
| SecondCell_RSRP | RSRP of the secondary cell | Float |
| SecondCell_SNR | SNR of the secondary cell | Float |
| SecondCell_PSC | PCI of the secondary cell | Integer |
| SecondCell_ARFCN | ARFCN of the secondary cell | Integer |
| NTech1 | Radio access technology of the first neighbouring cell (e.g., 4G, 5G) | String |
| NCellid1 | Cell ID of the first neighbouring cell | Integer |
| NLAC1 | Location Area Code of the first neighbouring cell | Integer |
| NCell1 | Node identifier of the first neighbouring cell | String |
| NARFCN1 | ARFCN of the first neighbouring cell | Integer |
| NRxLev1 | RSRP of the first neighbouring cell, typically -140 to -44 dBm | Float |
| NQual1 | RSRQ of the first neighbouring cell, typically -19.5 to -3 dB or missing if unavailable. | Float |
| PINGAVG | Average latency of ping tests to a server in milliseconds | Float |
| PINGMIN | Minimum latency observed during ping tests in milliseconds | Float |
| PINGMAX | Maximum latency observed during ping tests in milliseconds | Float |
| PINGSTDEV | Standard deviation of ping latencies in milliseconds | Float |
| PINGLOSS | Number of packets lost during ping tests, indicating reliability | Float |
| TESTDOWNLINK | Measured downlink speed during an FTP test in Mbps | Float |
| TESTUPLINK | Measured uplink speed during an FTP test in Mbps | Float |
| TESTDOWNLINKMAX | Maximum achievable downlink speed during an FTP test in Mbps | Float |
| TESTUPLINKMAX | Maximum achievable uplink speed during an FTP test in Mbps | Float |
| Test_Status | Status of the current test with values like upload, download, ping | String |
| Mobility | Mobility state of the device (e.g., Walking, Driving) for easy categorization. | String |
| Node_Longitude | Longitude of the serving base station | Float |
| Node_Latitude | Latitude of the serving base station | Float |
| SessionID | Unique identifier for a continuous measurement session per operator, starting from 0. | Integer |
| ElapsedTime | Time elapsed since the start of the session in seconds | Float |

## 4.1 Data Collection Methodology

The methodology for collecting and utilizing the data was divided into two main parts: (1) Data collection using *GnetTrack Pro,* and (2) Data Preprocessing. The final output is a dataset ready for analytics, machine learning, and deep learning applications. Fig 1 illustrates a data-driven workflow for real-time mobile network analysis and optimization. It begins with data collection via field tests using GNetTrack Pro, which uploads data to a cloud database. The raw data is then downloaded, merged and processed using Python scripts for cleaning and feature engineering. This structured data is used for training and testing machine learning/deep learning models, with further stages involving performance visualization and storage of results in a processed database. The pipeline supports intelligent analysis for tasks like mobility prediction and network quality assessment.

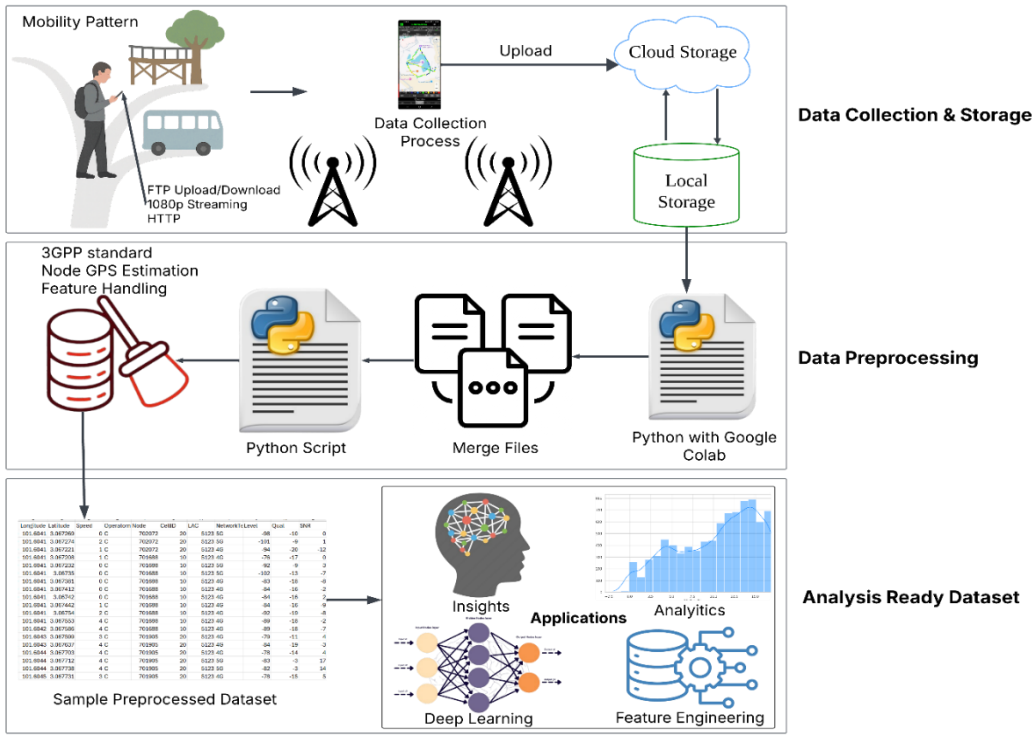

**Figure 1.** Data Collection Methodology

### 4.1.1 Data Collection

Data was collected with Samsung S21 Ultra (LTE Cat 20 DL / Cat 18 UL, up to 2 Gbps/200 Mbps), which is capable of measuring variants of 5G, using the *GNetTrack Pro* app. The dataset spans across three major operators in Malaysia, covering routes that include high-traffic pedestrian zones, canopy trails, BRT lines, and shuttle buses to reflect realistic user experiences. The collected session logs were uploaded to the cloud, and then subsequently merged using a Python script executed on Google Collaboratory, resulting in the final consolidated dataset. Fig 2 illustrates the coverage area considered, which is the publicly accessible outdoor areas within the 2km radius of Sunway city.

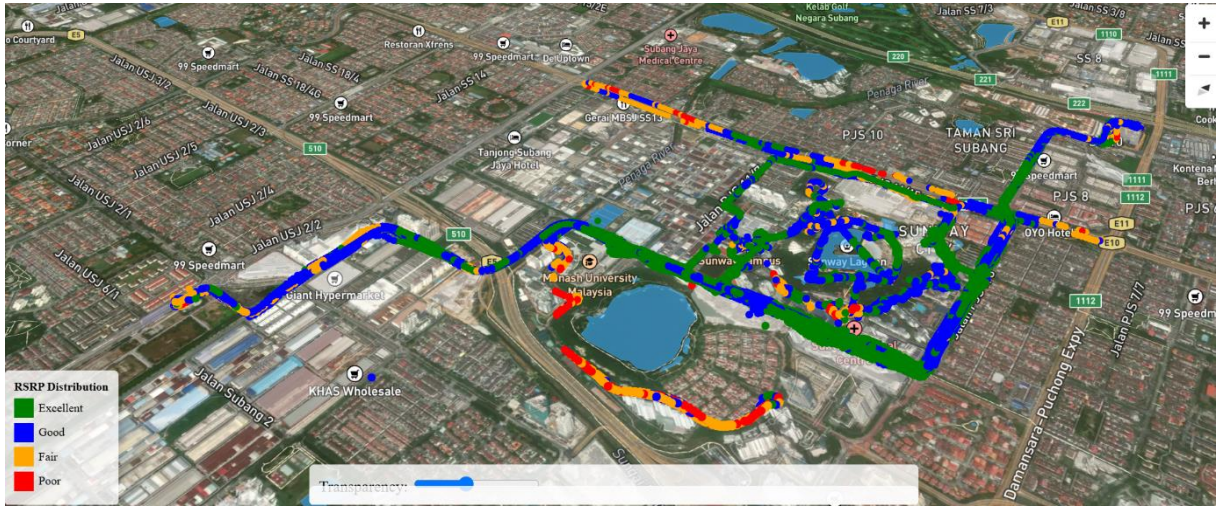

**Figure 2.** Map showing drive-test paths and RSRP distribution.

### 4.1.2 Data Preprocessing

The collected dataset is passed through a Python script data processing pipeline for handling outliers as per 3GPP TS 36.214 standard, handling missing values and outliers. The dataset has diversity for additional features to be engineered using appropriate method to produce the final analysis ready dataset, python scripts used for initial analysis and visualizations of the dataset is shared along with the dataset. Fig 3 shows a sample processed file, the operator names has been anonymized for privacy reasons.

**Figure 3.** Sample Processed file in .CSV Format

### 4.2 Collecting Data and Exploring Locations

A pilot data collection phase was conducted at the beginning, during which devices were calibrated, and measurement validation were performed. It was observed that some 5G devices are limited to Sub-6GHz measurements and do not support other variants like the mmWave. Based on these findings, the Samsung S21 Ultra was selected for its measurement versatility, and it was consistently used for data collection across all operators to avoid bias. Fig 4 shows some of the data collection sites, GnetTrack pro interface and some physical nodes verified.

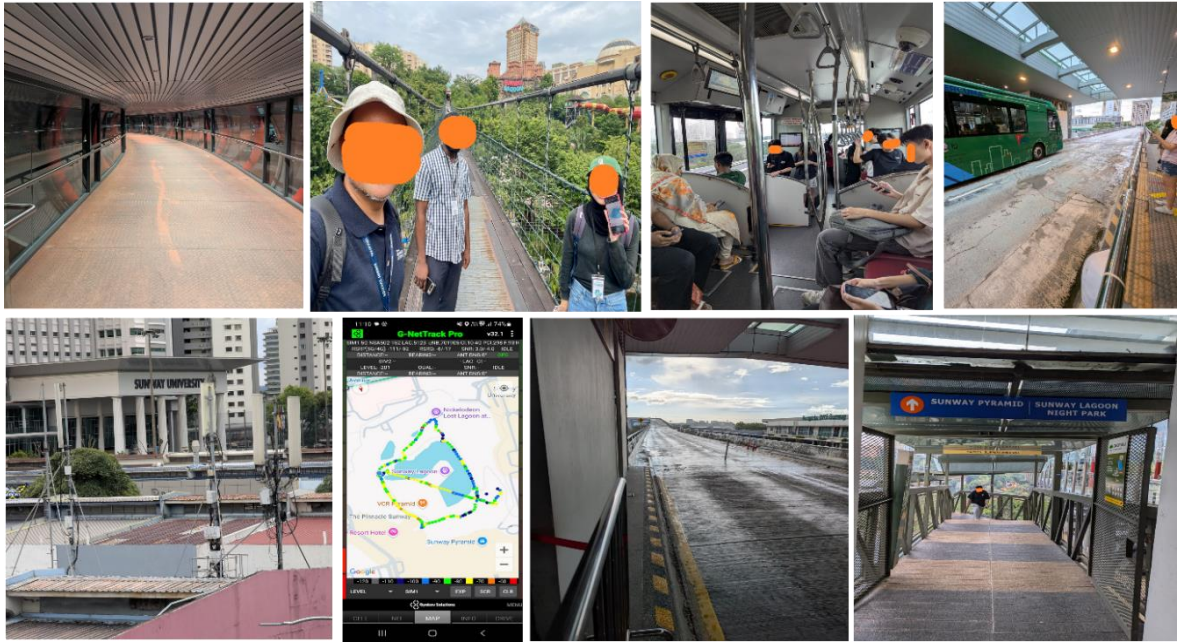

**Figure 4.** Sample locations and modes of data collection

## 4.3 Performed Experiments

The initial analysis performed on the collected dataset include the following: (1) Distribution Analysis and (2) Nodes Location Estimation

### 4.3.1 Distribution Analysis

Distribution analysis evaluates how the dataset is distributed across different operators and network technology, as well as in relation to mobility modes and traffic types.

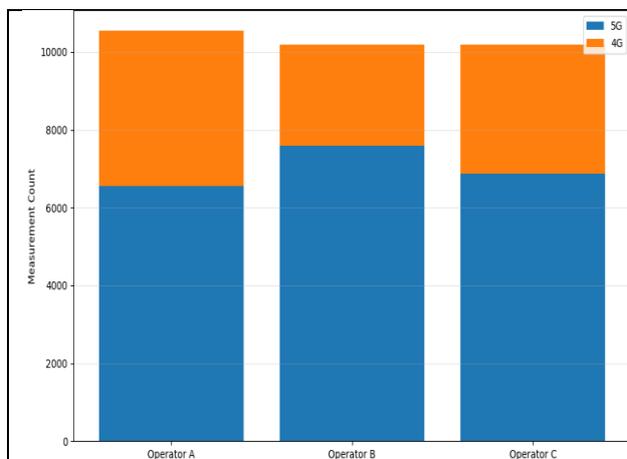

**Figure 5.** Network Technology Distribution per Operator

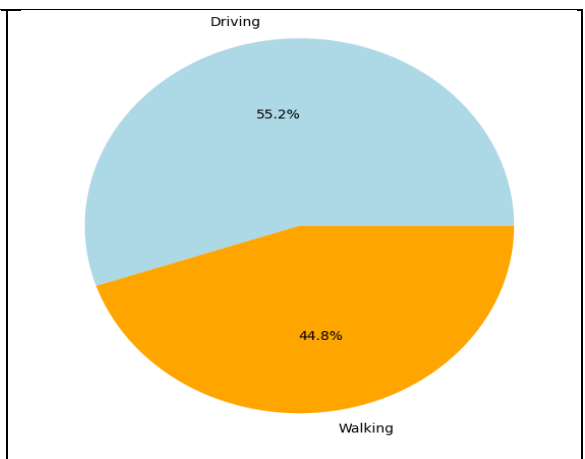

**Figure 6.** Distribution based on Mobility Pattern by Operator.

Fig 5 shows that Operator B has the most extensive 5G deployment, followed by Operators A and C, while 4G still maintains a significant presence across all three. The chart also indicates that 5G comprises approximately 70% of the dataset, confirming a non-uniform 5G rollout, which is typical of urban environments where factors such as infrastructure sharing, and operator strategy affect deployment. Figure 6 depicts the distribution of mobility patterns, categorized as walking and driving, with driving data forming the majority.

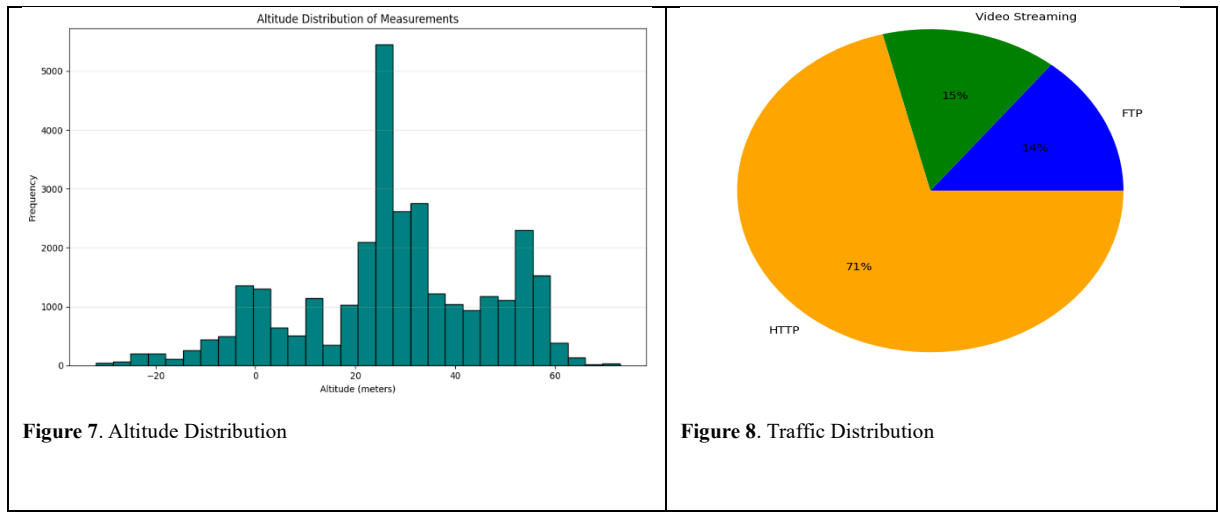

**Figure 7**. Altitude Distribution     **Figure 8**. Traffic Distribution

Fig 7 presents the altitude distribution, ranging from –32 to 73 meters with a mean of 27.42 meters. The diversity in elevation capturing canopy walkways, ground-level commutes, and areas near underpasses or basements enriches signal propagation analysis. This dimension is especially relevant in cities with vertical zoning, where altitude-aware coverage and beamforming are critical for reliable 5G service. Fig 8 aggregates traffic analysis across all operators and shows the distribution of FTP, Video Streaming and HTTP traffic, reinforcing the dataset's relevance in traffic-aware analysis. These findings are important for QoE analysis, from another perspective, the less represented classes can be augmented using Synthetic Minority Over-sampling Technique (SMOTE) or Generative Adversarial Networks (GAN) to generate more records.

### 4.3.2 Nodes Location Estimation

A centroid-based method was used to estimate the geographic coordinates of cell sites, based on multiple measurements and signal strength data collected via GNetTrack Pro. A total of 132 unique nodes were identified, with analysis revealing infrastructure sharing among some operators.

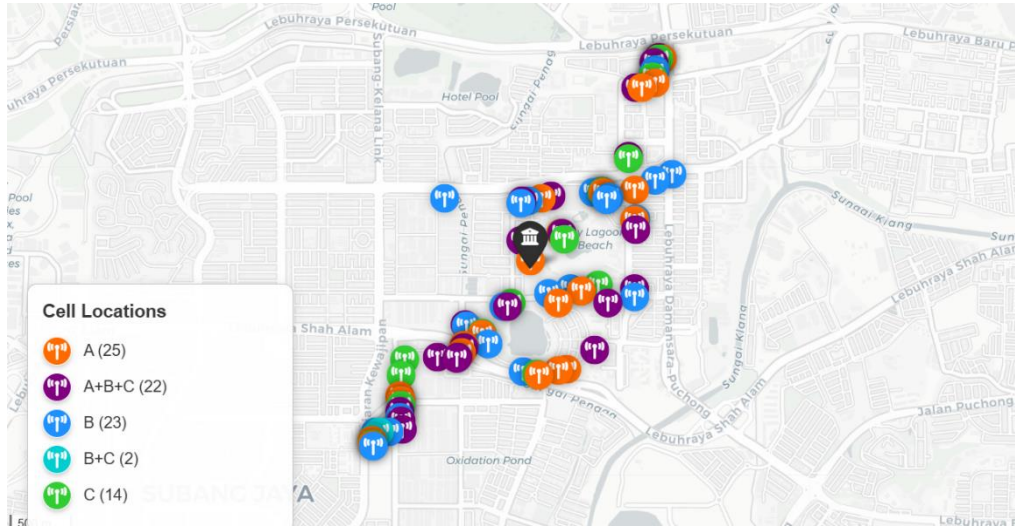

**Figure 9.** Estimated Cell Locations for 3 Operators including Shared Nodes

Fig 9 visualizes the identified physical cell sites, representing a typical high-density 5G urban deployment. Shared infrastructure is indicated, emphasizing the growing importance of active and passive sharing agreements in modern network rollouts.

The results of these preliminary analysis highlight the strength of the Sunway Drive test Dataset as a rich, mobility-sensitive, application-aware, and multi-operator dataset. By focusing on a high-density urban environment and incorporating altitude, mobility, and traffic profiles, this dataset fills a key gap left by synthetic or less context-aware urban datasets. It supports advanced research in areas such as handover prediction [20], QoE modelling [8], signal metrics regression [1], and infrastructure planning, all grounded in empirically validated, real-world data collected using standardized tools. This dataset is therefore not just a static snapshot of network conditions, it is a dynamic foundation for research and development in urban cellular networks.

**Limitations**

- Dataset is limited to a single urban location in Malaysia.
- Data collected only on Android devices using GNetTrack Pro.

**Ethics Statement**

No human subjects were involved. All data collected via sensors with no personal information, cellular network operator names are anonymized.

**Credit Author Statement**

Muhammad Kabeer, Rosdiadee Nordin, Mehran Behjati: Conceptualization, Methodology, Muhammad Kabeer, Farah Yasmin binti Mohd Shaharuddin: Route Planning, Data curation, Muhammad Kabeer: Writing Original Draft, Visualization, Investigation, Rosdiadee Nordin, Mehran Behjati: Project administration, Supervision, Muhammad Kabeer, Rosdiadee Nordin, Mehran Behjati: Reviewing and Editing.

**Data Availability**

The dataset is publicly available at [An Urban Multi-Operator QoE-Aware Dataset for Cellular Networks in Dense Environments](#) (Mendeley Data). It includes:

1. Raw CSV file with all records
2. Python script for preprocessing
3. Pre-processed CSV file

**Acknowledgements**

This work is supported by Sunway University with reference number GRTIN-RAG(02)-DEN-03-2024. We would like to also acknowledge Poh Kam (Final Year BSc Computer Science Student at Sunway) for help in data collection.

**Declaration of Competing Interest**

The authors declare that they have no known competing financial interests or personal relationships that could have appeared to influence the work reported in this paper.

**Declaration of Generative AI and AI-assisted Technologies in the Writing Process**

During the preparation of this manuscript, the authors used GPT-4o to enhance grammar and improve textual coherence.

**Reference**

[1] M. F. Ahmad Fauzi, R. Nordin, N. F. Abdullah, and H. A. H. Alobaidy, "Mobile Network Coverage Prediction Based on Supervised Machine Learning Algorithms," *IEEE Access*, vol. 10, pp. 55782–55793, 2022, doi: 10.1109/ACCESS.2022.3176619.

[2] V. Silva *et al.*, "A multi-device and multi-operator dataset from mobile network coverage on Android devices," *Data Brief*, vol. 57, Dec. 2024, doi: 10.1016/j.dib.2024.111146.

[3]     D. Raca, D. Leahy, C. J. Sreenan, and J. J. Quinlan, "Beyond throughput, the next generation: A 5G dataset with channel and context metrics," in *MMSys 2020 - Proceedings of the 2020 Multimedia Systems Conference*, Association for Computing Machinery, Inc, May 2020, pp. 303–308. doi: 10.1145/3339825.3394938.

[4]     A. U. Rehman, M. Bin Roslee, and T. Jun Jiat, "A Survey of Handover Management in Mobile HetNets: Current Challenges and Future Directions," Mar. 01, 2023, *MDPI*. doi: 10.3390/app13053367.

[5]     A. F. Mostafa, M. Abdel-Kader, and Y. Gadallah, "A UAV-based coverage gap detection and resolution in cellular networks: A machine-learning approach," *Comput Commun*, vol. 215, pp. 41–50, Feb. 2024, doi: 10.1016/j.comcom.2023.12.010.

[6]     M. Al-Khalidi, R. Al-Zaidi, N. Thomos, and M. J. Reed, "Intelligent Seamless Handover in Next-Generation Networks," *IEEE Transactions on Consumer Electronics*, vol. 70, no. 1, pp. 1566–1579, Feb. 2024, doi: 10.1109/TCE.2023.3340416.

[7]     K. Technologies, "Next-Generation Wireless: A Guide to the Fundamentals of 6G," 2023.

[8]     A. Haghrah, M. P. Abdollahi, H. Azarhava, and J. M. Niya, "A survey on the handover management in 5G-NR cellular networks: aspects, approaches and challenges," Dec. 01, 2023, *Springer Science and Business Media Deutschland GmbH*. doi: 10.1186/s13638-023-02261-4.

[9]     A. Mostafa, M. A. Elattar, and T. Ismail, "Downlink Throughput Prediction in LTE Cellular Networks Using Time Series Forecasting," in *2022 International Conference on Broadband Communications for Next Generation Networks and Multimedia Applications, CoBCom 2022*, Institute of Electrical and Electronics Engineers Inc., 2022. doi: 10.1109/CoBCom55489.2022.9880654.

[10]    M. F. A. Fauzi, R. Nordin, N. F. Abdullah, H. A. H. Alobaidy, and M. Behjati, "Machine Learning-Based Online Coverage Estimator (MLOE): Advancing Mobile Network Planning and Optimization," *IEEE Access*, vol. 11, pp. 3096–3109, 2023, doi: 10.1109/ACCESS.2023.3234566.

[11]    M. Behjati, M. A. Zulkifley, H. A. H. Alobaidy, R. Nordin, and N. F. Abdullah, "Reliable Aerial Mobile Communications with RSRP & RSRQ Prediction Models for the Internet of Drones: A Machine Learning Approach," *Sensors*, vol. 22, no. 15, Aug. 2022, doi: 10.3390/s22155522.

[12]    H. A. Alobaidy *et al.*, "Empirical 3D Channel Modeling for Cellular-Connected UAVs: A Triple-Layer Machine Learning Approach," 2025, arXiv preprint arXiv:2505.19478.

[13]    S. Verma, S. Abhirami, A. Kumar, and S. D. Amuru, "Double Deep Reinforcement Learning assisted Handovers in 5G and Beyond Cellular Networks," in *2023 15th International Conference on COMmunication Systems and NETworkS, COMSNETS 2023*, Institute of Electrical and Electronics Engineers Inc., 2023, pp. 466–470. doi: 10.1109/COMSNETS56262.2023.10041356.

[14]    Y. Jang, S. M. Raza, H. Choo, and M. Kim, "UAVs Handover Decision using Deep Reinforcement Learning," in *Proceedings of the 2022 16th International Conference on Ubiquitous Information Management and Communication, IMCOM 2022*, Institute of

Electrical and Electronics Engineers Inc., 2022. doi: 10.1109/IMCOM53663.2022.9721627.

[15] T. H. Sulaiman and H. S. Al-Raweshidy, "Predictive handover mechanism for seamless mobility in 5G and beyond networks," *IET Communications*, vol. 19, no. 1, Jan. 2025, doi: 10.1049/cmu2.12878.

[16] C. K. Nayak, S. Karunakaran, P. Yamunaa, S. Kayalvili, M. Tiwari, and M. V. Unni, "Analysis of Digital Twins Implementation in Smart City using Big Data and Deep Learning," in *Proceedings of the 7th International Conference on Intelligent Computing and Control Systems, ICICCS 2023*, Institute of Electrical and Electronics Engineers Inc., 2023, pp. 230–235. doi: 10.1109/ICICCS56967.2023.10142813.

[17] S. K. Biswash, "Device and network driven cellular networks architecture and mobility management technique for fog computing-based mobile communication system," *Journal of Network and Computer Applications*, vol. 200, Apr. 2022, doi: 10.1016/j.jnca.2021.103317.

[18] S. M. Shahid, J. H. Na, and S. Kwon, "Incorporating Mobility Prediction in Handover Procedure for Frequent-handover Mitigation in Small-Cell Networks," *IEEE Trans Netw Sci Eng*, 2024, doi: 10.1109/TNSE.2024.3487415.

[19] E. Selvamanju and V. B. Shalini, "Deep Learning based Mobile Traffic Flow Prediction Model in 5G Cellular Networks," in *3rd International Conference on Smart Electronics and Communication, ICOSEC 2022 - Proceedings*, Institute of Electrical and Electronics Engineers Inc., 2022, pp. 1349–1353. doi: 10.1109/ICOSEC54921.2022.9952113.

[20] A. Tahat, F. Wahhab, and T. A. Edwan, "An Exemplification of Decisions of Machine Learning Classifiers to Predict Handover in a 5G/4G/3G Cellular Communications Network," in *20th International Conference on the Design of Reliable Communication Networks, DRCN 2024*, Institute of Electrical and Electronics Engineers Inc., 2024. doi: 10.1109/DRCN60692.2024.10539173.